\documentclass[twocolumn,showpacs,preprintnumbers,amsmath,amssymb,floatfix,superscriptaddress,nofootinbib,prl]{revtex4-1}

\usepackage{graphicx}
\usepackage{subfig}
\usepackage{epsfig}
\usepackage{amsfonts}
\usepackage{amssymb}
\usepackage{epsf}
\usepackage{color}
\usepackage{xcolor}
\usepackage{stmaryrd}
\usepackage[utf8]{inputenc}

\usepackage{tabularx} 
\usepackage{amsmath}  
\usepackage[final]{hyperref} 
\hypersetup{
	colorlinks=true,       
	linkcolor=blue,        
	citecolor=blue,        
	filecolor=magenta,     
	urlcolor=blue         
	}

\newcommand{\gb}{\mathcal{G}}

\newcommand{\meff}{m_{\text{eff}}}
\newcommand{\beq}{\begin{equation}}
\newcommand{\eeq}{\end{equation}}

\begin{document}


\title{Compact object scalarization with general relativity as a cosmic attractor}
 

\author{{\bf Georgios Antoniou}}
\affiliation{School of Mathematical Sciences, University of Nottingham, University Park, Nottingham, NG7 2RD, UK}

\author{{\bf Lorenzo Bordin}}
\affiliation{School of Physics and Astronomy, University of Nottingham, University Park, Nottingham, NG7 2RD, UK}

\author{{\bf Thomas P. Sotiriou}}
\affiliation{School of Mathematical Sciences, University of Nottingham, University Park, Nottingham, NG7 2RD, UK}
\affiliation{School of Physics and Astronomy, University of Nottingham, University Park, Nottingham, NG7 2RD, UK}

\begin{abstract}
We demonstrate that there are theories that exhibit spontaneous scalarization in the strong gravity regime while having General Relativity with a constant scalar as a cosmological attractor. We identify the minimal model that has this property and discuss  its extensions. 
\end{abstract}

\maketitle

At the time of writing, there are 13 confirmed detections of compact object mergers via their gravitational wave emission \cite{LIGOScientific:2018mvr,Abbott:2020uma,LIGOScientific:2020stg}. This number is expected to rise to the hundreds in the coming years and this will allow us to probe the structure of neutron stars and black holes to unprecedented accuracy. We can then confront the predictions of General Relativity (GR) with observation and test the theory itself in the strong gravity regime. However, one might ask: is it reasonable to expect significant deviations from GR in the strong field regime, considering that the theory has been tested to extremely high precision in the weak field?

 \emph{Spontaneous  scalarization} is perhaps the most direct manifestation of new physics that stays dormant in the weak field regime and yet leads to large deviation from GR in the strong field regime. The first model that exhibits the spontaneous scalarization phenomenon was proposed by Damour and Esposito-Far\`ese (DEF) in \cite{Damour:1993hw}. 
Here, a direct coupling between a scalar field $\phi$ and the Ricci scalar, $R$ (or equivalently the matter in a different conformal frame), generates at linear level an effective mass for $\phi$. As the compactness of an objects increases, this effective mass can become negative and trigger a tachyonic instability. The scalar field then grows until nonlinear effects kick in and quench the instability, thereby leading to a ``scalarized'' object: a neutron star that is dressed with a scalar configuration and, hence, has different structure than its GR counterpart.\footnote{In the DEF model this is simply due to the backreaction of the scalar and the effect this has on the star, but in principle one could have additional couplings to matter that alter the microphysics within the star as well \cite{Coates:2016ktu,Franchini:2017zzx}.} Since the effective mass is proportional to $R$ in the DEF model, no instability can be triggered around 
black holes that are solution of GR.\footnote{A subtle exception are black hole that have matter configurations in their vicinity \cite{Cardoso:2013fwa,Cardoso:2013opa}.} In fact, the DEF model is covered by no-hair theorems \cite{Bekenstein:1995un, Hawking:1972qk, Sotiriou:2011dz}. 

A coupling between the scalar and the Gauss-Bonnet invariant 
$\gb \equiv R^2 - R_{\mu\nu}R^{\mu\nu} + R_{\mu\nu\rho\sigma}R^{\mu\nu\rho\sigma}$ 
has been known to evade no-hair theorems and lead to scalar hair \cite{Campbell:1991kz,Mignemi:1992nt,Kanti:1995vq,Yunes:2011we,Sotiriou:2013qea,%
Sotiriou:2014pfa,Antoniou:2017acq,Doneva:2017bvd,Silva:2017uqg,Antoniou:2017hxj}.
Recently, it has been shown that, in models that fashion such a coupling but also admit GR solutions with constant $\phi$, black hole and neutron star scalarization can occur \cite{Silva:2017uqg, Doneva:2017bvd}.  Hence, scalarization is not specific to neutron stars or to the DEF model. Further investigations have demonstrated that the properties of the scalarized object are sensitive to nonlinear interactions \cite{Blazquez-Salcedo:2018jnn,Silva:2018qhn,Macedo:2019sem}. The onset of scalarization is instead controlled only by interaction terms that contribute to linear perturbations around a GR background, as scalarization commences as a linear tachyonic instability. 

Schematically, linearizing around a GR solution and neglecting backreaction, one has
 \begin{equation}
\Box_{\rm eff}\,\phi^{(1)} -  \meff^2[\alpha^i,\phi^{(0)},g^{(0)}_{\mu\nu}]\,\phi^{(1)} +{\rm NL}=0.
\end{equation}
where $g^{(0)}_{\mu\nu}$ is the GR background with $\phi^{(0)}=$constant, $\phi^{(1)}$ is the linear scalar perturbation, $\alpha^i$ collectively denotes the coupling constants of the theory, and ${\rm NL}$ stand for nonlinear interactions that can be neglected at linear order. $\Box_{\rm eff}$ is the d'Alembertian of either $g^{(0)}_{\mu\nu}$ or  some effective metric and $\meff^2$ can be seen as an effective mass squared, whose value is controlled by the coupling constant but also the background. Hence, when the coupling constants and the background satisfy certain conditions, $\meff^2$  can become sufficiently negative and the scalar undergoes a tachyonic instability, as mentioned earlier in the context of the DEF model. The nonlinear terms cease to be negligible, quench the instability and determine its endpoint. One can follow this reasoning and pin down the most general set of terms that will contribute to $\Box_{\rm eff}$ and $\meff^2$ and thereby to the onset of the instability in scalar-tensor theories \cite{Andreou:2019ikc}. The mechanism could be generalized to non-gravitational couplings \cite{Herdeiro:2018wub} and other fields \cite{Ramazanoglu:2017xbl,Ramazanoglu:2018hwk}. 

Spontaneous scalarization models rely on the fact that $\meff^2$ depends on curvature. This allows for objects characterized by high curvature to scalarize, while objects charactirized by low curvature will be described by GR solutions with $\phi=\phi^{(0)}$. There is a thorny subtlety though: if one treats these objects as isolated and hence asymptotically flat, as usual, then one can always assume that  $\phi=\phi^{(0)}$ asymptotically. However, in a more realistic setup the value of $\phi$ far away from the object is actually determined by cosmological considerations. As it turns out, when the coupling constant of the DEF model is such that scalarization can occur for neutron stars, GR solutions with $\phi=\phi^{(0)}$ are {\em not} attractors in late time cosmology \cite{Damour:1992kf}, see also \cite{Anderson:2016aoi} for a more recent detailed analysis. Similarly, models that exhibit black hole scalarization due to a coupling between the scalar and the Gauss-Bonnet invariant also exhibit exponential growth of the scalar during cosmological constant domination \cite{Franchini:2019npi}. Hence, without severely fine-tuning initial conditions in cosmology, localized matter configurations in the late universe could not be described by GR with $\phi=\phi^{(0)}$ and scalarization models would be effectively ruled out. 

The aim of this paper is to instead point out that (generalized) scalar-tensor theories that have GR as a cosmological attractor and still exhibit scalarization at large curvatures actually exist. We will first demonstrate this by means of a simple (perhaps the simplest) example and argue intuitively why this is expected. We will then proceed to discuss the cosmology of such models a bit more thoroughly, discuss how generic our results are and explain how they would change in more general classes of scalarization models. 

Let us consider the following action,
\begin{align}
\label{eq:theory}
S  =  & \frac{1}{2\kappa}  \int   d^4x \sqrt{-g}\left[
R + 4X - 2\beta \phi^2 R + 2\lambda L^2 \ \phi^2\gb \right]
\end{align}
 where $X=-(\partial\phi)^2/2$ is the kinetic term of the scalar field, $\beta$, $\lambda$ are coupling constants and $L$ is an additional lengthscale that one needs to choose. We assume that the metric is minimally coupled to matter.
 The corresponding scalar equation of motion is 
\begin{equation}\label{sc}
\Box\phi + ( \lambda L^2 \gb-\beta R)\phi =0.
\end{equation}
The couplings with $R$ and ${\cal G}$ generate an effective mass for the scalar field, 
\begin{equation}\label{eq:eff_mass}
\meff^2 = \beta R-\lambda L^2  \gb\,.
\end{equation}
We are interested in models that exhibit spontaneous scalarization around compact objects so we need to demand that $\meff^2$ becomes negative at high curvature in order to trigger a tachyonic instability. For the time being our goal is to just demonstrate that this simple model can exhibit spontaneous scalarization for some type of compact objects and still have GR as a cosmological attractor. So, we restrict attention to spherical black holes. Our GR solution will then be the Schwarzschild solution, for which we have $R=0$ and 
$\gb = 12 r_s^2/r^6$.
As mentioned earlier, $\gb$ is then sign-definite and the condition for having a negative $\meff^2$ becomes $\lambda > 0$ \cite{Silva:2017uqg}. For scalarization to be relevant to astrophysical black holes we need to choose $L$ to be of the order of the characteristic lengthscale of the compact object, so we choose $L\sim 10 \, {\rm km}$. Finally, we stress that for GR solutions to be admissible in the model under consideration, one should have $\phi=\phi^{(0)}=0$. Hence, this is the asymptotic value that $\phi$ would need to take for unscalarized configurations.

Next, we turn our attention to studying this theory on cosmological scales. Assuming 
a flat Friedman-Lema\^itre-Robertson-Walker metric, 
the equation of motion for $\phi$ is,
\begin{equation}
\label{eq:cosmo_scalar_eom}
\ddot \phi + 3H\dot \phi + \meff^2(t)\phi=0\,,
\end{equation}
where $\meff^2(t)$ is given by eq.~\eqref{eq:eff_mass} and depends on the cosmological background.
To get the evolution of the scale factor $a(t)$ we study the $tt$ component of the modified Einstein Equations
\begin{equation} \label{eq:tt_ee}
G_{tt} = \, \kappa \, \left(\rho_\phi + \rho_a \right)\,,
\end{equation}
where  $\rho_a$ denotes the energy densities of the various conventional components of the cosmic fluid and $\rho_\phi$ is an {\em effective}  energy density associated with the scalar field, given by
\begin{equation}\label{eq:sc_density}
\rho_\phi = \kappa^{-1} \left[ \dot\phi^2 +6\beta H^2 \phi^2 +12 H \phi \dot\phi \left( \beta - 4 \lambda L^2 H^2 \right) \right]\,.
\end{equation} 
The cosmic fluid is well approximated by a barotropic fluid whose pressure is given by $p_a = w_a \rho_a$, with the index $a=r,m, de$ and $w_a=1/3,\,-1,\,0$ for radiation domination (RD), matter domination (MD) and dark energy domination (DED) respectively.

We do not require $\phi$ to play any role in late universe cosmology, so we will assume that it is subdominant with respect to $\rho_a$.
This assumption helps avoid the gravitational wave constrains on Dark Energy theories (see \cite{Creminelli:2017sry,Ezquiaga:2017ekz,Sakstein:2017xjx,Baker:2017hug,Creminelli:2018xsv}), as discussed in detail in \cite{Franchini:2019npi}. Under the condition $\rho_\phi \ll \rho_a$
Eq.~\eqref{eq:tt_ee} simplifies to the usual Friedmann equation, $H^2\approx \kappa \rho_a/3$.
This, together with the continuity equation, $\dot{\rho}_a + 3H\rho_a(1+w_a) = 0$ allows us to simplify the expressions for the curvature terms
\begin{align}
    R = & \ 6(2H^2+\dot H) = \kappa \, \rho_a\,(1-3{w_a}), \\
    \gb = & \ 24H^2 (H^2+\dot H) = -\frac{4}{3} (\kappa \, \rho_a)^2(1+3w_a),
\end{align}
and hence the expression for the effective mass.

Let us now return to  \eqref{eq:cosmo_scalar_eom} and consider the behaviour of the scalar in different cosmological eras. Table \ref{tab:signs} summarizes the signs of the Ricci scalar, $R$, and the Gauss-Bonnet invariant, ${\cal G}$, during each era. Note that these, together with the signs of the coupling constants $\beta$ and $\lambda$, control the sign of the effective mass. It is also worth emphasising that $R$ and ${\cal G}$ have different dimensions and hence different scaling with time, with ${\cal G}$ being clearly dominant at earlier times. 

\begin{table}[htb]
\centering
\begin{tabular}{c c c c}
& \hspace{3mm}Radiation\hspace{3mm} & \hspace{3mm}Matter\hspace{3mm} & \hspace{3mm}Dark Energy\hspace{3mm} \\
\hline\hline
$\gb$ & $<0$ & $<0$ & $>0$ \\
$R$ & $0$ & $>0$ & $>0$ \\
\hline
\end{tabular}
\caption{\footnotesize Signs of the Ricci scalar and the Gauss-Bonnet invariant during  different cosmological eras. \label{tab:signs}}
\end{table}

During RD, $R$ effectively vanishes and, hence, the mass of the scalar field is entirely controlled by the ${\cal G}$ term, with $\meff^2 \simeq - \lambda L^2 \gb \approx 24 \lambda H^4 L^2 \propto 1/t^4$, since $H\propto 1/t$. 
At very early times $\meff^2$ will dominate over the friction term in Eq.~\eqref{eq:cosmo_scalar_eom}. However, $\meff^2$ decays much faster than the Hubble friction and the latter will rapidly take over and drive $\phi$ to a constant.
The time that $\phi$ takes to freeze is approximatively given by the time at which the potential is comparable with the Hubble friction. 
After this point, it only takes a few Hubble times for $\dot\phi$ to effectively vanish. More concretely, 
 $\meff \lesssim H \Rightarrow H(z) \times L \lesssim 1$, which happens very early, around the redshift $z\approx10^{11}$ for our choice of $L$. As a result, the scalar field is already frozen to a constant solution well before MD. 
 
 At the onset of MD, $\phi$ starts evolving again. 
This is because $R$ no longer vanishes on cosmological scales and thus it provides a non-negligible contribution to $\meff^2$. 
The contribution of the ${\cal G}$ term in $\meff^2$ has actually become largely subdominant to that of the $R$ term of their different scaling. 
During MD, $H\ll L^{-1}$.

As has been pointed out in \cite{Andreou:2019ikc}, action \eqref{eq:theory} with $\lambda=0$ is related by a simple field redefinition to a linearized version of the DEF model. In fact, we have defined $\beta$ such that $8\beta=\beta_{\rm DEF}$ in the appropriate limit. Nonlinearities are not important in our regime. As a result, one expects that once the ${\cal G}$ term in our theory has become negligible, cosmological evolution will match that of the DEF model. Interestingly, the latter actually exhibits our desired cosmological behaviour for $\beta>0$ \cite{Damour:1992kf}: GR is a cosmological attractor! Hence the scalar field will naturally be driven to $\phi=0$. 
The transition to DED does not chance the dynamics of the scalar qualitatively and GR with $\phi=0$ continues to be the attractor.

All of the above can be verified by studying the scalar dynamics quantitatively. 
In this regard it is better to express Eq.~\eqref{eq:cosmo_scalar_eom} in terms the redshift, in which case it takes the following form:
\beq\label{eq:cosmo_scalar_eom2}
\phi_{(a)}''+f_a\phi_{(a)}'+q_a\phi_{(a)}=0,
\eeq
where prime denotes differentiation with respect to $z$, with
\begin{align}
f_a(z)=&\frac{H'(z)}{H(z)}-\frac{2}{z+1},\\
q_a(z)=&\frac{12 L^2 H(z)^2 (\lambda +3 \lambda  w_a)+3 \beta (3 w_a-1)}{(z+1)^2}.
\end{align}
We begin our numerical analysis at $z_{i} = 10^{10}$,  just before Big Bang Nucleosynthesis (BBN). 
To set the initial conditions for the scalar field and its derivative, we assume that $\phi$ is just coupled with the thermal bath.
Therefore a natural initial value is $\phi_{i} \simeq H(z_{i}) / \kappa \ll 1$.  
The initial value $\phi'_{i}$ can be, instead, derived from $\dot \phi_{in}$: we expect $\dot \phi_{i} \simeq H(z_{i}) \phi_{i} \Rightarrow \phi'_{i} \simeq \phi_i / z_{i}$, which is, again, much smaller than unity.
These two conditions ensure that $\rho_{\phi}(z_{i}) \ll \rho_r(z_{i})$ and are hence consistent with the assumption that $\phi$ is cosmologically subdominant. 
We stress that $\phi \sim 1$ would imply Planckian energy scales in our units and hence initial conditions with $\phi_{i} \ll 1$ do not constitute fine tuning.

\begin{figure*}[t]
\begin{tabular}{cc}
\includegraphics[width=0.9\textwidth]{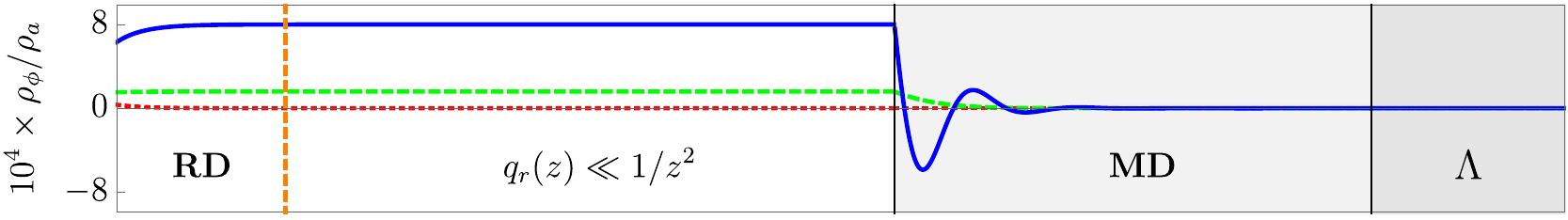} & \\[-2mm]
\includegraphics[width=0.9\textwidth]{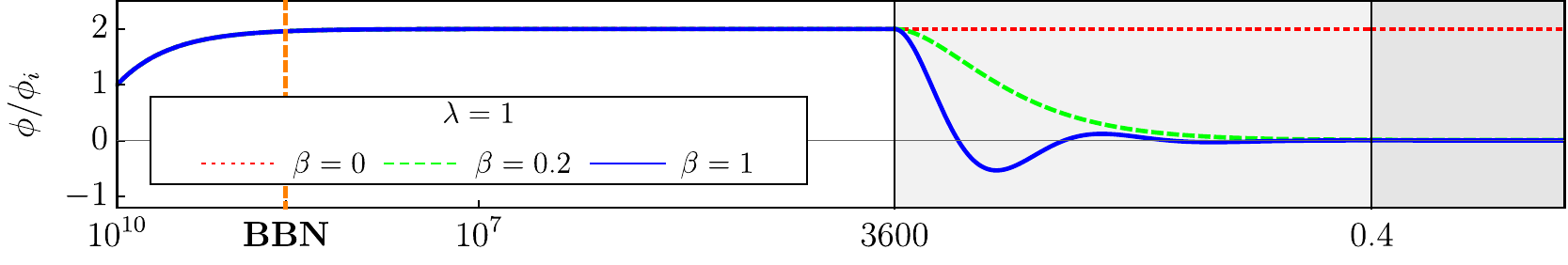} & \\
\end{tabular}
\caption{\label{fig:fig1} {\em Top panel:} Effective energy density of the scalar $\rho_\phi$ over  the energy density of the cosmic fluid $\rho_a$ as a function of redshift. {\em Bottom panel:} Evolution of the scalar field $\phi$ in units of its a reference value $\phi_i$, fixed at $z=10^{10}$.}
\end{figure*}

Fig.~\ref{fig:fig1} shows the evolution of the scalar and of the ratio $\rho^\phi/\rho_a$
for $z<z_i$. $\rho_\phi$ remains subdominant  as expected and the plots confirm the qualitative behaviour described previously. In particular, $\phi$ remains constant throughout, with the exception of transitions between cosmological eras.  

The value $\phi$ takes at late times does depend crucially on $\beta$. For $\beta=0$, $\phi$ effectively remains frozen to the value it has in the early RD era. Unless this value is set to be extremely close to zero by fine tuning initial data, any local configuration in the late universe will have to be scalarized because cosmological asymptotics will be incompatible with having unscalarized configurations. As discussed in the introduction, this would clash with weak field constraints. For $\beta>0$ instead, $\phi\to 0$ during MD and GR with $\phi=0$ becomes a cosmological attractor. To approach this attractor fast enough, $\beta$ should be of order unity so that the oscillations seen in Fig.~\ref{fig:fig1}
at the onset of MD are nearly \emph{critically damped}. 
These oscillations correspond to changes on the effective Newton's constant  that will, in principle, affect the formation of Large Scale Structures. 
However, the time scale of the oscillations is very large, of order of the Hubble rate. Moreover, the corrections to Newton's constant would be $\propto |\beta| \, \Delta \phi^2$, and hence negligible.
In summary, cosmic evolution is expected to be almost identical to GR for late times.

\begin{figure}[t]
\begin{tabular}{c}
\includegraphics[width=0.4\textwidth]{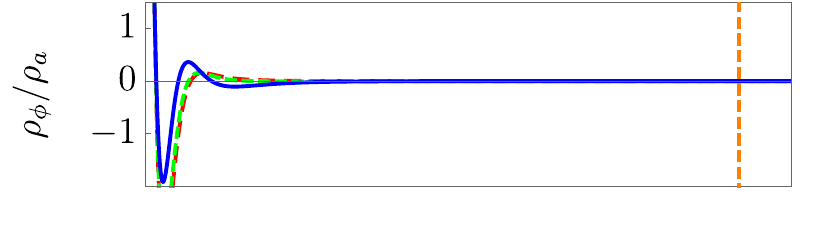} \\[-7mm]
\includegraphics[width=0.4\textwidth]{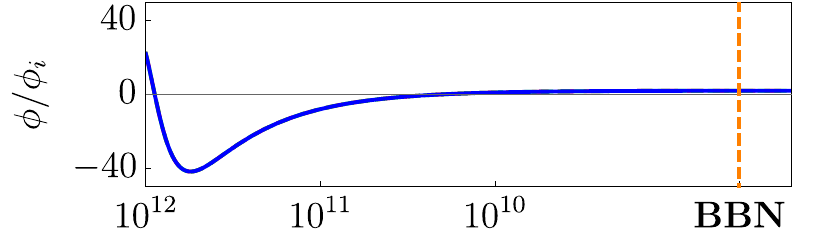}
\end{tabular}
\caption{\label{fig:fig3} Same as Fig.~\ref{fig:fig1} but for very high redshifts.}
\end{figure}

Fig.~\ref{fig:fig3} shows the evolution of $\rho_\phi$ and $\phi$ for $z>z_i$ and the very early epochs before recombination. As anticipated in our qualitative analysis, for $z \gg 10^{11}$, the significant contribution of the ${\cal G}$ term to the effective mass results in a sinusoidal behaviour. The oscillation is damped by Hubble friction when we move forward in time. $\rho_\phi$ also shows oscillatory behaviour and, moving to higher redshift, the oscillations are amplified. Eventually, our approximation that $\phi$ is subdominant ceases to be valid. It is worth emphasising that $\rho_\phi$ does not need to remain positive, as it is just an effective energy density.

\begin{figure}[t]
\begin{tabular}{cc}
\includegraphics[width=0.4\textwidth]{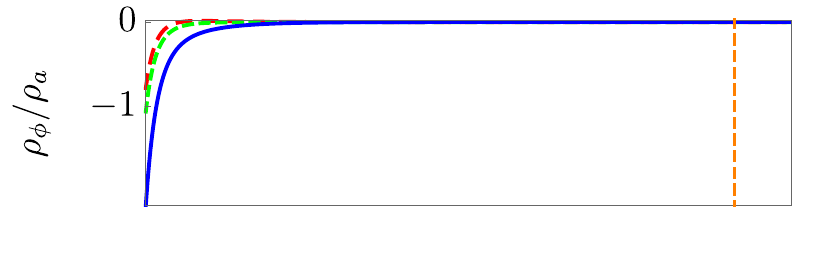} & \\[-8mm]
\includegraphics[width=0.4\textwidth]{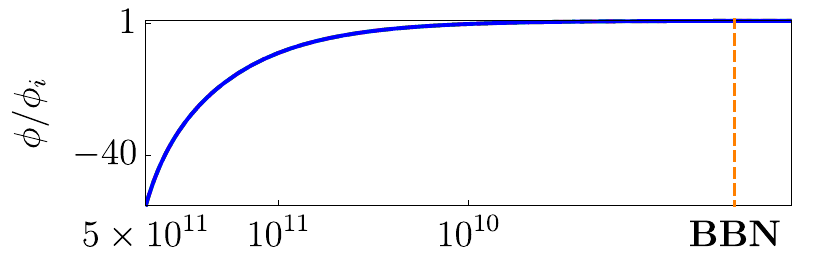} & \\
\end{tabular}
\caption{\label{fig:fig4} Same as Fig.~\ref{fig:fig3} but for $\lambda=-1$.}
\end{figure}

All of the above referred to $\lambda, \beta>0$. Next we discuss the case $\beta>0$, $\lambda<0$. 
On astrophysical scales, $\lambda < 0$ leads to spontaneous scalarization triggered by a tachyonic instability in the interior of neutron stars \cite{Silva:2017uqg}. 
As the previous analysis has already shown, on cosmological scales the $\lambda$ term has an impact only at very early times, before $z\simeq10^{11}$. Indeed, numerical analysis confirms that flipping the sign of $\lambda$ makes no difference during BBN and at later times. However, as seen from Table \ref{tab:signs}, for $\lambda<0$, the $\lambda {\cal G}$ contribution to $\meff^2$ will be negative and will lead to exponential growth of $\phi$ once one reaches sufficiently $z$ for the mass contribution to dominate over Hubble friction. 
As shown in Fig.~\ref{fig:fig4}, $\rho_\phi/\rho_a$ grows exponentially fast and reaches $1$ a lot earlier than when $\lambda>0$. 

Note that,
since scalarization relies on curvature couplings, it is rather intuitive that the terms that trigger it will become relevant in the very early universe. The coupling with the Gauss-Bonnet invariant is the dominant one at large curvatures and its coupling constant is dimensionful. As such, it controls the curvature scale at which departure from standard cosmology would appear. This would happen when the universe is of the size of a few kilometres, well before BBN, for values of the coupling that are compatible with compact object scalarization. At earlier times, departures from standard cosmology would be significant, as our results show, and as has been pointed out in the literature \cite{Anson:2019uto}. However, it is quite a stretch to consider these models as good effective field theories, and hence take their predictions seriously, all the way to energy scales where the universe is the size of kilometres. Instead, it seems sensible to try to embed them in a suitable UV completion with suitable inflationary cosmology. 

Finally, we consider $\beta<0$. For $\lambda=0$, one expects to recover the results of Refs.~\cite{Anderson:2016aoi}. In fact, for any value of $\lambda$ one will have a tachyonic instability  on cosmological scales at late times. This instability will be very slow, so it is not particularly threatening in its own right. However, without an attractor mechanism at late times, severe tuning of initial conditions would be needed to have GR configurations locally (see $\beta=0$ case) and the instability would only make things worse.

To conclude, we have demonstrated, using a specific model as an example, that the phenomenon of spontaneous scalarization around compact objects is compatible with having 
 an attractor mechanism to GR on cosmological scales. In fact, our result show that fairly simple scalarization models can track GR cosmology over a vast range of redshift and all the way back to BBN. The key feature that leads to the desired behaviour is that the scalar can couple in two different ways to curvature --- through the Gauss-Bonnet invariant and through the Ricci scalar --- with one coupling triggering scalarization locally and the other providing a late time attractor cosmologically.
 
The action we have considered is rather minimal, as it only includes terms that contribute to linearized perturbations around GR solutions with constant scalar. It is perfectly sufficient to discuss the onset of scalarization and whether GR is cosmological attractor. However, the properties of scalarized solutions will be controlled by the nonlinear (self)interactions of the scalar that one can add to our action \cite{Silva:2018qhn,Macedo:2019sem,Andreou:2019ikc}. Hence, there is actually a wide variety of scalarization models with the desired cosmological behaviour at late time and different properties for compact objects. We leave the study of more elaborate models and the properties of compact objects in such models for future work.

 \begin{acknowledgments}
{\bf \em Acknowledgements:} L.B. and T. P. S. acknowledge partial support from the STFC Consolidated Grant No. ST/P000703/1. G.A. acknowledges partial support from the Onassis Foundation. We would also like to acknowledge networking support by the COST Action GWverse CA16104.
 \end{acknowledgments}
 
\bibliography{biblio}

\end{document}